\documentclass[10pt, conference]{IEEEtran}

\ifCLASSINFOpdf
   \usepackage[pdftex]{graphicx}
  \graphicspath{{../pdf/}{../jpeg/}{../png/}}
  \DeclareGraphicsExtensions{.pdf,.jpeg,.png}
\else
\fi
\usepackage{url}

\begin{document}
\title{Keeping Continuous Deliveries Safe}


\author{
\IEEEauthorblockN{Sebastian V\"ost}
\IEEEauthorblockA{Research and Development\\
BMW Group\\
Munich, Germany\\
sebastian.voest@bmw.de}
\and
\IEEEauthorblockN{Stefan Wagner}
\IEEEauthorblockA{Institute for Software Engineering\\
University of Stuttgart\\
Stuttgart, Germany\\
stefan.wagner@informatik.uni-stuttgart.de}

}


%Telephone: (800) 555--1212, Fax: (888) 555--1212}


\maketitle

\begin{abstract}
Allowing swift release cycles, Continuous Delivery has become popular in application software development and is starting to be applied in safety-critical domains such as the automotive industry.

These domains require thorough analysis regarding safety constraints, which can be achieved by formal verification and the execution of safety tests resulting from a safety analysis on the product. With continuous delivery in place, such tests need to be executed with every build to ensure the latest software still fulfills all safety requirements. Even more though, the safety analysis has to be updated with every change to ensure the safety test suite is still up-to-date.

We thus propose that a safety analysis should be treated no differently from other deliverables such as source-code and dependencies, formulate guidelines on how to achieve this and advert areas where future research is needed. 

\end{abstract}


% \fi
\IEEEpeerreviewmaketitle

\section{Introduction}

Agile software development practices have seen a major boom in the past decades due to their ability in coping with changing requirements and enabling short release cycles. 
This has climaxed in today's movements of DevOps and continuous software engineering \cite{bosch2014continuous}. The latter aims to integrate every aspect of software development, starting from sourcecode and code analysis up to the delivery to productive environments, in a \textit{Continuous Integration Pipeline}. The ultimate goal is being able to deliver the latest software to the customer at the push of a button. 

In corporate environments, other aspects have to be considered to the extent that even processes such as finance and HR are sometimes examined from a continuous point of view \cite{fitzgerald2014continuous}.

Indeed, traditional corporate environments are adapting to use agile software practices. With the fourth industrial revolution, many industrial branches that were traditionally alien to computers are now facing a huge part of their products being driven by software. A key example for this is the automotive industry, which is challenged by the need to mass-produce cyber-physical systems like no other industrial branch.

At present, efforts are being made to widely establish such a pipeline for software integration in the industry \cite{Vost2015}, while one manufacturer appears to be having such a pipeline in place already. Recent accidents indicate, however, that safety is a key feature to be taken into account more than ever with the appearance of autonomous cars.

In fact, safety is arguably the most significant difference between cyber-physical systems and application software. Other aspects, such as organizational corporate matters or service availability are issues for application and mobile developers as well.

\subsection{Problem Statement}

Since agile and continuous software development is transcending to cyber-physical-systems, it is crucial to investigate how safety requirements can be integrated in these concepts. Nevertheless, while there are efforts to include safety engineering into agile software processes \cite{wang2016toward} \cite{staalhane2012application}, as of now (October 2016) there is no research on continuous software engineering in combination with safety.

\subsection{Hypothesis}
We assume that, while formal verification on the design is possible, it is always  based on abstractions, an additional step of safety testing on the concrete product is necessary.

Thus, our hypothesis is that an advanced continuous delivery pipeline for safety-critical systems should include the safety analysis as an artifact such as source code, dependencies or tests and integrate test cases generated from the analysis into every build. After all, the goal of a continuous delivery pipeline is to deploy changed code into the production environment at the push of a button. Thus, this is the only way to ensure that all safety requirements are met.

\subsection{Contributions}

In this paper, we present our idea of \textit{Continuous Safety Builds} and what is required to implement them, outline the \textit{Implications} of such a process for programmers and team members, propose \textit{Possible Tool Support} along with \textit{Required Research} and finally highlight the \textit{Expected Benefits} from an integrated safety analysis. Related work is presented inline where appropriate.

\section{Continuous Safety Builds}

\begin{figure}[h!]
\centering
\includegraphics[width=3.5in]{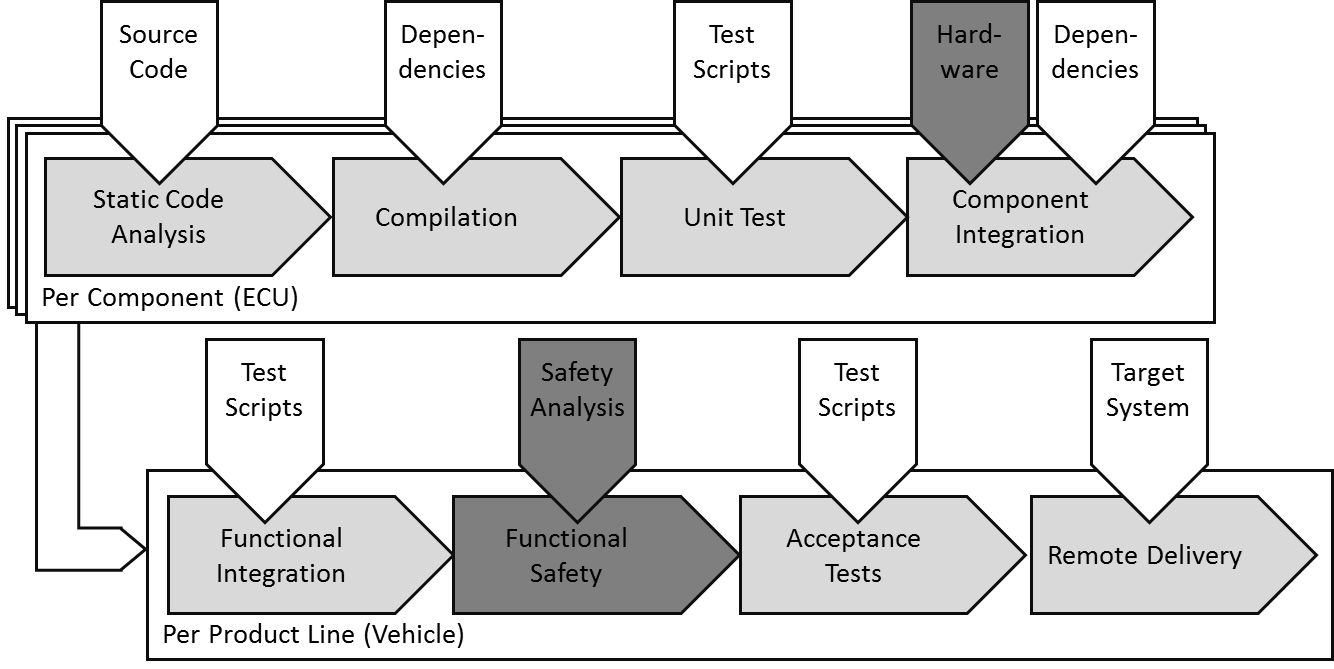}
\caption{Deployment Pipeline in the Automotive Industry}
\label{fig_cd_pipeline}
\end{figure}

The delivery pipeline in the automotive domain is quite similar to its counterpart in application software development, as depicted by an example in Figure \ref{fig_cd_pipeline}. The major differences are highlighted in dark grey: 

Every component of the system is integrated on the target hardware before a functional integration can take place. This is necessary, because modern complex functions are distributed over several self-contained control units that need to communicate to implement a function on the target system, the vehicle. 

The second major difference are the safety tests that are performed on the system, and this one sticks out. For all other steps, there are well-established or experimental implementations that allow integration into a Continuous Delivery Suite \cite{Vost2015} \cite{stolberg2009enabling} \cite{de2005generic}, but not for this one.

All steps in the pipeline use the result of the previous step and combine them with one or more artifacts, mainly automated test scripts. Other examples for artifacts may be rules for static code analysis, hardware or third-party libraries. All of them have in common, that they are developed parallel to the source code, kept up-to-date on central infrastructure (such as a CI system or source repository) and that an update in any of these should cause a new build.

Based upon this notion, we formulate the following guidelines for Continuous Safety Builds:

\textbf{1. Iterative Safety Analysis performed in parallel to development}

An up-to-date safety analysis is the first prerequisite for contiunous safety testing. The analysis must be maintained iteratively in line with the source code. This idea has been introduced by Stahlhane et al. \cite{staalhane2012application} with \textit{Safe Scrum}, that integrates safety analysis into Scrum sprints.

Analysis methods are often designed to be iterative with respect to a changing system design or architecture \cite{bozzano2003esacs} \cite{cassanelli2006failure}.

\textbf{2. Safety Test execution and generation needs to be automated}

Automated execution of safety test cases is an obvious requirement, but the test cases need to be automatically generated from the analysis as well. Since the safety analysis needs to be integrated into development, the necessary manual work needs to be reduced to a minimum.

\begin{figure}[t!]
\centering
\includegraphics[width=3.5in]{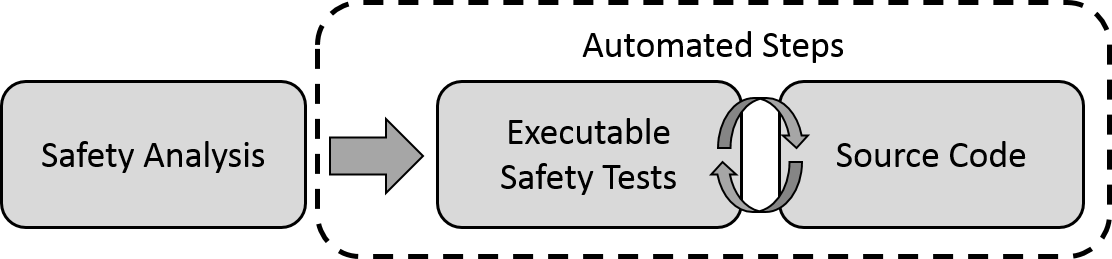}
\caption{Steps in Testing Software against Safety Analysis}
\label{fig_automation_steps}
\end{figure}

As depicted in Figure \ref{fig_automation_steps}, the safety analyis has to be transformed into concrete test cases that can be executed on the software. This step is often done manually, but there are approaches to automate this process fully, such as the XSTAMPP platform \cite{abdulkhaleq2015comprehensive} for STPA \cite{leveson2013stpa}.

\textbf{3. The Safety Analysis is an artifact like the source code or test scripts}

The Safety Analysis and its results should be treated in no way differently from other artifacts required for a build. Source code, tests and configuration are often stored in the same repository and dependencies are downloaded from servers that keep them up-to-date. The same should be true for the safety analysis, it needs to be stored in a central repository and in a versioned manner. 

If this holds, every change in the analysis can be tracked and even trigger a build, which includes the generation of tests from the analysis and the execution of updated tests against the latest program version. This allows to apply the benefits of continuous integration to the process of safety engineering and shows inconsistencies immediately and automatically.

\textbf{4. Safety Tests are included in every build}

If the above is given, we can achieve our ultimate goal, which is to check every latest piece of software for compliance with the safety analysis. Thus, every build includes the safety tests generated from the analysis. 

This ensures that every build which might eventually be delivered to production has passed an automatic safety check. If this check was not passed, the build breaks and thus we ensure that potentially unsafe software is not deployed even in a CD scenario.

\section{Implications}

\subsection{Seamless Integration of Safety and Software Engineering}
To continuously integrate new software with the latest results from the safety analysis, we need to ensure that we can always test against an up-to-date safety analysis. Thus, performing this analysis cannot be a separate process but must be closely tied to the software development.

In Safe Scrum \cite{staalhane2012application}, every sprint in a scrum process is expanded by a safety backlog in addition to the standard sprint backlog, which includes the safety-relevant issues that are changed in a single iteration.

For a continuous pipeline, however, this is still not sufficient, because the work items from both backlogs might be scheduled differently. This may lead to a part of safety-relevant software being updated and commited, while the safety analysis is not up-to-date (or vice versa) and thus breaking the build.

\begin{figure}[h!]
\centering
\includegraphics[width=3.5in]{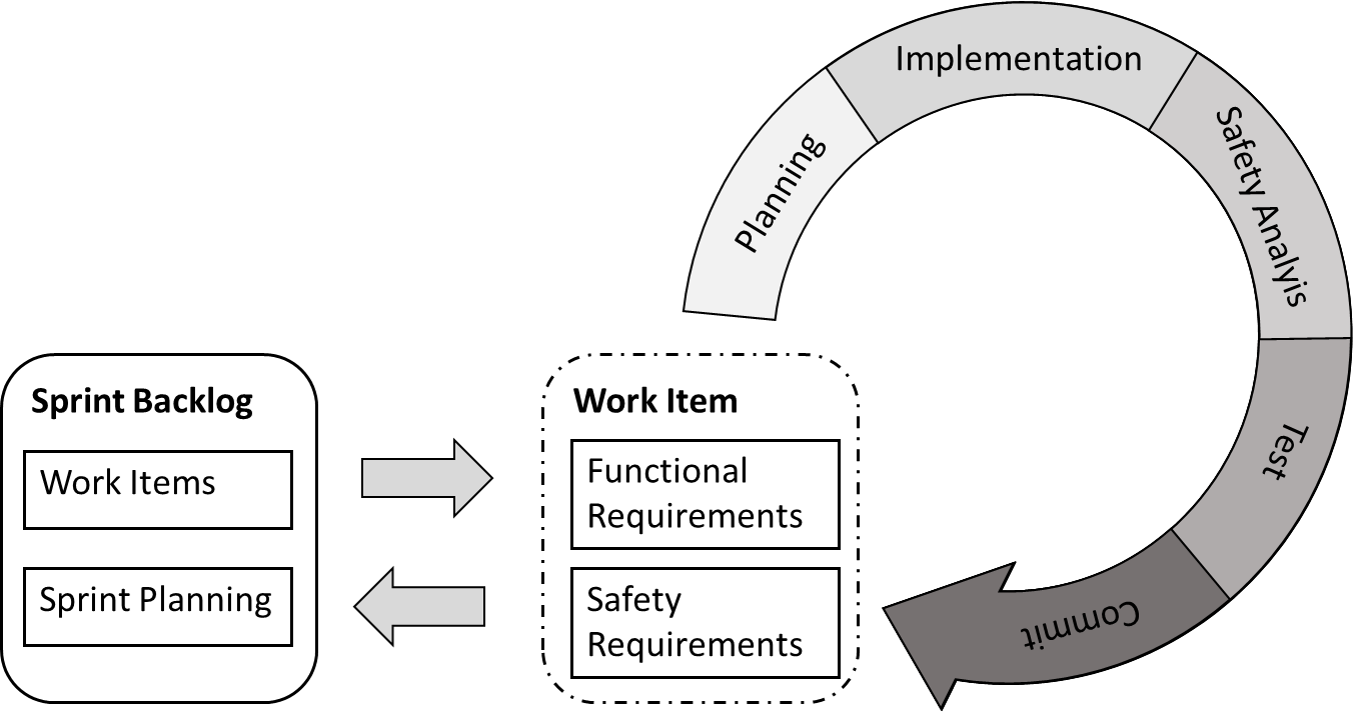}
\caption{Adapted workflow for Work Items during a sprint}
\label{fig_dev_process}
\end{figure}

Instead, the safety backlog needs to be integrated with the functional sprint backlog. As depicted in Figure \ref{fig_dev_process}, each work item is enriched with the safety requirements affected by it and updating the safety analysis if necessary becomes a required step before a commit to the mainline is possible.

\subsection{Every Developer becomes a safety engineer}
Since we propose that tests generated from the safety analysis are able to break the build, this requires an additional check before commiting to the mainline and thus triggering a build. At the very least, a developer has to be aware whether or not his commit affects any of the safety-relevant components and changes their behaviour. 
If it does, they need to ensure that the safety analysis is updated together with the changed source code in order to keep the build running. If the changes are small, they might even be able to make the adjustments themselves. 

\subsection{Dedicated Safety Engineers are supported by developers, not replaced}
Including developers in the process of safety engineering does not mean that dedicated safety engineers become obsolete. The necessity to view safety from a system perspective still holds, but developers will only view the aspects of the module they are working on. 

A safety engineer will thus be required to maintain and expand the safety analysis for every release and keep it in line with the requirements of the entire system. However, they will be able to rely on the developers' groundwork and aggregate their changes to a reworked analysis.

\section{Possible Tool Support}

An implementation of the proposed process requires adaptations of currently available tools. In the following section, we focus on Continuous Delivery suites, integrated developer environments and tools supporting safety analysts and give ideas how such tools could be expanded.

\subsection{Integrated Developer Environments}

Since we propose to involve developers in safety analysis tasks and want to intertwine them with coding, it is straightforward to directly offer developers support in the tools they use in their daily business.

Popular integrated developer environments (IDEs) such as IntelliJ \cite{intellij} or Visual Studio \cite{visualstudio} are highly configurable by plugins. Plugins for developers of safety-critical software could create traceability between the source-code and the safety analyis. Safety-critical parts of the software could be marked, e.g. by annotations or models for traceability between design and safety constraints \cite{briand2014traceability} could be connected to the implementation.  

That way, the IDE could point the developer to the relevant parts of the safety analysis and/or possibly affected safety requirements, if a safety-relevant change was made to allow him to make modifications with ease if necessary. 

More concretely, in an example where STPA \cite{leveson2013stpa} is used, the IDE could present a list of possibly affected hazards and unsafe control actions for the developer to review and change if necessary.

Updated safety artifacts could be commited to the repository along with the source code without breaking a developer's workflow. This enables easy adoption of guideline 1 stated in chapter II. 

%\begin{itemize}
%\item plugins uses safety analysis in a certain format, downloaded from repository
%\item Connection between sourcecode and safety analysis is established. possibly by annotations, maybe take into account the system architecture...
%\item Checkup if changes in the sourcecode affect the safety analysis
%\item Reminder/List/View: "Have your changes affected these safety requirements?", allow to mark as irrelevant, but point the developer to the parts of the safety analysis that need to be modified
%\end{itemize}

\subsection{Safety Analysis Tools}

Safety analysis methods typically feature tools that support a safety engineer in doing so. Some of them, such as the XSTAMPP platform \cite{abdulkhaleq2015comprehensive} for STPA, already allow partial or full generation from safety test cases when a certain step in the analysis is reached. 

%This is the level of automatization that is required for our process. The user provided some artifacts, in the example of XSTAMPP that would be the STAMP model, and the system control structure. The tool performs all further steps and generates executable test cases.

As stated above, traceability to the source code is desirable. Thus, such tools should be expanded to allow connection to the system design and implementation. 

Finally, tools should make sure that their deliverables are stored in an open, machine-readable and non-binary format to allow exchange between different tools and to track changes in a version control system.

\subsection{Continuous Delivery Suites}

Continuous Delivery Suites such as GitLab \cite{gitlab} support developers by integrating a source management system with a customizable delivery pipeline. Each stage of the pipeline can be configured to call external tools or execute scripts. 

Safety Analysis tools could be called to generate safety tests and the execution of tests could be triggered without much adaptation this way. The safety test cases would be executed along with all other test cases such that implementation of guidelines 3 and 4 can be achieved.

For further ease of use and support of a dedicated safety engineer, plugins for these tools could provide an user interface to aggregate the safety-specific tests and their latest results along with the recent changes in the safety analysis.

%\begin{itemize}
%\item Plugins that allow the upload of a Safety analysis in a certain format
%\item Test generation is automatically started upon change
%\item Tool supports switching between methods and combination of multiple
%\item Safety Coverage is measured
%\item notification of safety engineer if a relevant change has been identified
%\end{itemize}

\section{Required Research}

The main goal of this paper is to provide an incentive for researchers to look into related topics. At the present state of research, an implementation of the proposed process seems hard, because we do not know enough about some underlying requirements.

The most basic requirement is the automatic generation of safety test cases from well-defined deliverables of a safety analyst. While there are approaches and tools to do this, the full pipeline from a changed safety requirement to tests has yet to be automated. The viability of an integration of safety analysis into CI/CD suites also needs to be proven.

We have stated in the previous chapter that we would like to improve traceability between safety constraints and related source code, while we presented only very basic solutions. It is likely that more sophisticated methods like data mining algorithms or recommenders using historical data perform well in deciding whether or not a change in some lines of code has an effect on safety constraints. This should be further investigated.

Finally, as of now we have only formulated a hypothesis and proposed a process. The feasibility has yet to be shown in empirical research, where appropriate tool support is available and first experiences with Continuous Safety Builds are made.

\section{Expected Benefits and Impacts}

With this paper, we have presented a new aspect of safety engineering. Not only is our goal to integrate safety aspects into agile software development, but into a continuous integration and delivery process. Further research in this area will greatly help us as researchers to understand how producers of safety-critical systems can set up and maintain fast delivery cycles. At the same time, researchers can help such industries to find solutions to challenges they encounter in doing so.

We have given an incentive to researchers of safety analysis methods to put more emphasis on process automatization and thus to reduce effort and ultimately cost of safety testing by reasoning to intertwine safety engineering and development. 

Finally, we aim to integrate safety into the field of continuous software engineering and thus to support continuous deliveries of functionally tested, but also safe software.

\section{Risks}

Talking about benefits, risks of a newly proposed method should also be adressed. A major issue is that a part of the safety engineering is assigned to the developers, who might not be appropriately educated to evaluate safety effects or make light-hearted decisions based on a good guess. This can be mitigated by rising the awareness of the importance of safety aspects in developers, providing appropriate tools, training and the support of a dedicated safety engineer.

Another obvious risk is that with the additional effort put into safety analysis, the productivity of a project may decline or the costs increase. Again, proper tool support can mitigate this. It is also possible that increased cost can be amortized by a lower liability for safety issues and/or savings due to the automatization of a previously manual process, but this would have to be shown in a larger study.

At last it is possible that managers and developers put too much faith into the Continuous Safety Builds and a full safety analysis is relinquished, which might lead to software being delivered that was not sufficiently tested in regard to safety, especially if above mentioned risks also arrive. This can only be mitigated by the awareness that a full re-evaluation of the safety analysis after a certain time interval cannot be replaced. However, this risk holds even more if safety-critical software is delivered without safety analysis and testing integrated into the delivery pipeline.

\section{Conclusion}

Investigating the delivery pipelines of safety-critical software, we have identified one major speciality: Safety testing, which is normally supported by a manual analyis using methods such as STPA. With regard to the current transition seen in consumer-oriented industries such as automotive software towards continuous delivery, we formulated the hypothesis that safety testing needs to be integrated into a continuous delivery pipeline.

We proposed that in order to achieve this, the safety analysis should be treated as an artifact no different from the source code itself and tests should be automatically generated from it, which can then be integrated in continuous builds. This implies a seamless integration of safety engineering into the development process. Every developer would have to iteratively update the safety analysis with each work item to prevent breaking safety builds, thus involving every team member in safety-critical tasks. 

Finally, we provided ideas for tool support in IDEs, Analysis Tools and Continuous Delivery Suites and pointed towards further research that is required to implement Continuous Safety Builds.

\bibliographystyle{IEEEtran}

\bibliography{library}
%\begin{thebibliography}{1}


%\end{thebibliography}

\end{document}